\documentclass[3p,twocolumn,nokeys]{elsarticle}

\usepackage{graphicx}
\usepackage{amssymb}
\usepackage{natbib}
\usepackage{lineno}

\makeatletter
\def\ps@pprintTitle{%
  \let\@oddhead\@empty
  \let\@evenhead\@empty
  \def\@oddfoot{\reset@font\hfil\thepage\hfil}
  \let\@evenfoot\@oddfoot
}
\makeatother

\begin{document}

\begin{frontmatter}

\title{Correction of Optical Aberrations in Elliptic Neutron Guides}

\author[a,b]{Phillip M. Bentley\corref{cor1}}
\ead{phil.m.bentley@gmail.com}

\author[a]{Shane J. Kennedy}
\author[b]{Ken H. Andersen}
\author[c]{Dami\'an Martin Rodr\'iguez}
\author[d]{David F. R. Mildner}

\cortext[cor1]{Corresponding Author}

\address[a]{Australian Nuclear Science and Technology Organisation (ANSTO), Locked Bag 2001, Kirrawee DC NSW 2232, Australia}
\address[b]{European Spallation Source, Stora Algatan 4, 223 50 Lund, Sweden}
\address[c]{J\"{u}lich Centre for Neutron Science, Forschungszentrum J\"{u}lich GmbH
52425 J\"{u}lich, Germany}
\address[d]{National Institute of Standards and Technology, Gaithersburg MD 20899, USA}

\begin{abstract}
Modern, nonlinear ballistic neutron guides are an attractive concept in neutron beam delivery and instrumentation, because they offer increased performance over straight or linearly tapered guides.  However, like other ballistic geometries they have the potential to create significantly non-trivial instrumental resolution functions.  We address the source of the most prominent optical aberration, namely coma, and we show that for extended sources the off-axis rays have a different focal length from on-axis rays, leading to multiple reflections in the guide system.  We illustrate how the interplay between coma, sources of finite size, and mirrors with non-perfect reflectivity can therefore conspire to produce uneven distributions in the neutron beam divergence, the source of complicated resolution functions.  To solve these problems, we propose a hybrid elliptic-parabolic guide geometry.  Using this new kind of neutron guide shape, it is possible to condition the neutron beam and remove almost all of the aberrations, whilst providing the same performance in beam current as a standard elliptic neutron guide.  We highlight the positive implications for a number of neutron scattering instrument types that this new shape can bring.
\end{abstract}

%
%


\end{frontmatter}


\hyphenation{}

\section{Introduction\label{sec:Introduction}}
Modern neutron guide systems increasingly feature a ballistic geometry, where the term ``ballistic'' refers to the middle section of the guide being wider than the guide entrance and exit.  This widening reduces the number of reflections required to transport a neutron beam compared to a straight neutron guide of the same length, but it also introduces optical changes that add complexities that have only recently become apparent.

The first ballistic guides featured linearly tapered geometry for the converging and diverging sections \cite{MEZEI_BALLISTIC_GUIDES,ABELE_BALLISTIC_TEST}.  Since this initial step, there has been progress \cite{SCHANZER_BALLISTIC} highlighting the benefits of switching to curved surfaces (or polygonal approximations to curves) following conic section geometries that have been exploited heavily in photon optics.

There are three types of conic section, depending on the properties of the beam and the desired result.  For a perfectly implemented optical system, a \emph{collimated} beam from an extended source can be focused onto a small target using a single reflection from a \emph{parabolic} mirror.

On the other hand, \emph{divergent} rays from a point source can be reflected by a single \emph{elliptic} mirror from one focal point to another focal point with one reflection.  This makes elliptic neutron guide shapes very attractive, as neutron beams are generally divergent at the source and we wish to minimise the number of reflections as far as possible to increase transport efficiency.

The third type of conic section is a \emph{hyperbolic} shape that brings \emph{convergent} rays to a nearer focal point.  At grazing angles and large distances from the focal point, linearly tapered guides are good approximations to hyperbolic geometries because hyperbolic curves asymptotically approach straight lines.

For point sources or targets, consider the inversion of the the optical system --- i.e. interchanging the target for the source --- it is clear that the properties of an ellipse are symmetric.   For the other two mirror geometries, a parabolic mirror would reflect a divergent beam from a point source and produce parallel rays; and a hyperbolic mirror would reflect a divergent beam such that it appears to be radiating from a farther focal point.

While reflectivity is the essential issue for the development of neutron guides, optical problems can arise when trying to use conic sections in neutron optics.  These can often be attributed to overlooking one or more of the following in the design:
\begin{enumerate}
\item Multiple reflections reduce the beam transport efficiency \label{cause:reflectivityCurve}
\item Neutron sources have finite spatial extent and cannot be treated as point sources\label{cause:PointSource}
\item No single type of conic section mirror deals with all incoming trajectories effectively\label{cause:MagicBullet}
\item The optimal focal point of the geometry does not necessarily overlap with the intended source or target\label{cause:FocalPoint}
\end{enumerate}

For example, it is very easy to design a ballistic guide system where the emerging beams have multi-modal divergence distributions at the sample position, which is the subject of this article.  The root cause of this essentially includes all of the above, and this leads to a non-trivial resolution function, which could be problematic in some experiments.

In many cases, making simple changes to the operating conditions and geometry to take these principles into proper consideration can lead to a large improvement in the beam characteristics.  This is an important issue, as elliptic guides are becoming very popular, almost to the extent that they are discussed in the context of being a ``magic bullet'' that is deployed to solve neutron transport problems where a different mirror geometry is probably more appropriate.

\begin{figure}
\includegraphics[width=70mm]{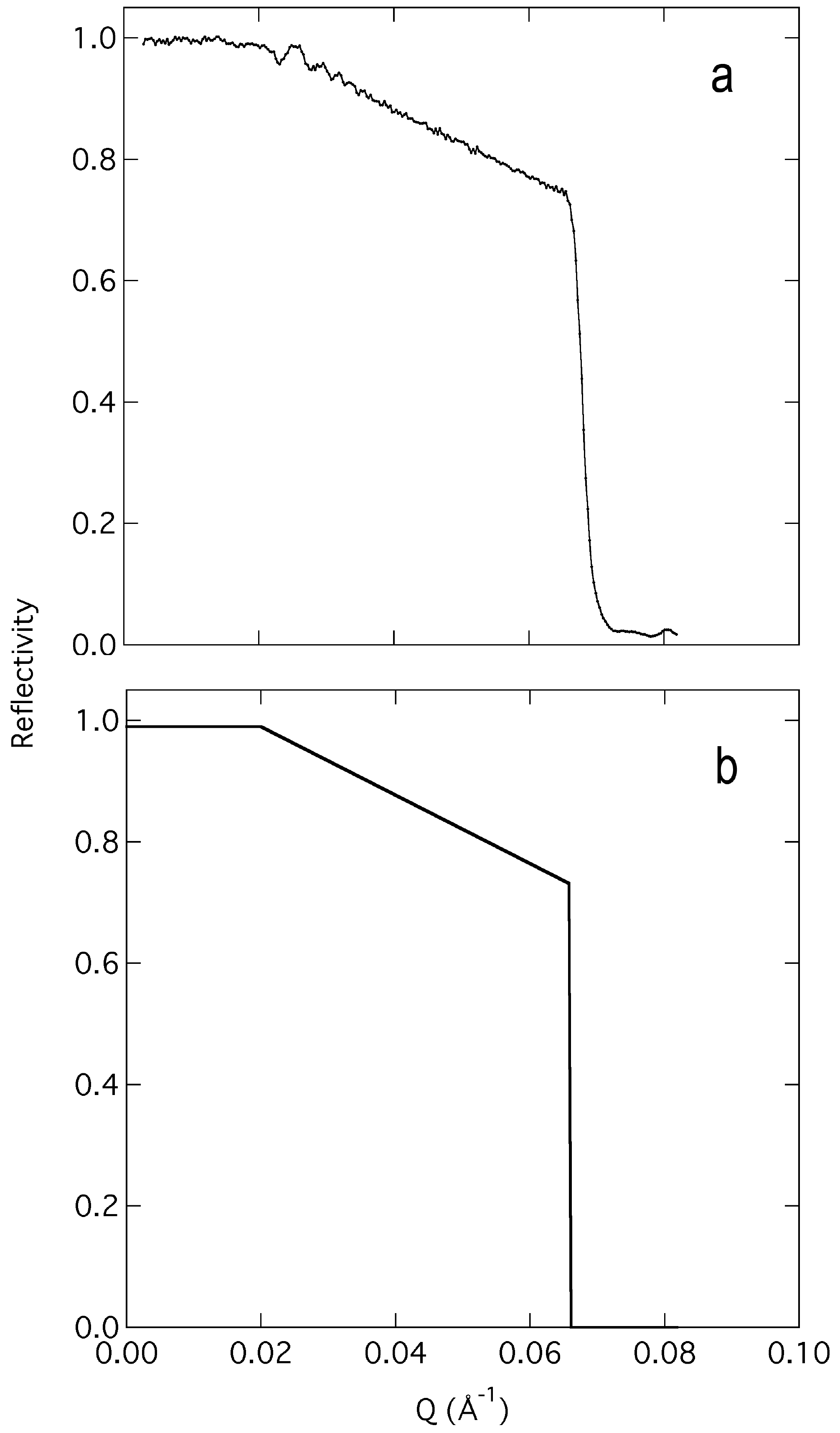}
\caption{(a) measured reflectivity of an $m=3$ supermirror as a function of neutron momentum transfer, $Q$ and (b) simplified model of the same supermirror using the same approximations that appear in NADS, after Bentley \& Andersen \cite{BENTLEY_NADS}.  Note that the critical reflectivity is rather poor for this particular mirror, which we use for illustration purposes, and that modern supermirrors offer much higher critical reflectivity.}
\label{fig:reflectivityCurves}
\end{figure}

We illustrate how optical aberrations arise in elliptic neutron guides, and how to eliminate them.  The beam modelling calculations have been performed using the established Monte-Carlo neutron ray tracing package VITESS \cite{VITESS} and a relatively new analytic method called ``neutron acceptance diagram shading'' (NADS) \cite{BENTLEY_NADS}.  These are two different approaches.  Aside from the statistical \emph{vs} analytic difference, NADS necessarily uses idealised, piecewise reflectivity curves with a sharp cut-off at the critical angle for reflection in the supermirrors (shown in figure \ref{fig:reflectivityCurves}).
We define the magnitude of the neutron momentum transfer vector $Q$ ($=4\pi \sin \theta / \lambda$), and $m$ is the critical momentum transfer for neutron reflection of the supermirror relative to that of natural nickel $Q_{Ni} = 0.0217$ \AA$^{-1}$, so that the critical reflectivity of the supermirror is $Q_{crit}=mQ_{Ni}$.  In the simplified model, the reflectivity decreases uniformly between $Q_{Ni}$ and $mQ_{Ni}$ according to:
\begin{equation}
\label{eq:NADSreflectivity}
R(Q)=\left\{
\begin{array}{ll}
 R_{Ni} & 0\leq Q \leq Q_{Ni} \\
 
R_{Ni}-g(Q-Q_{Ni})  &
   Q_{Ni}<Q\leq m Q_{Ni}  \\
   
   0 & Q > m Q_{Ni}
   
\end{array}
\right.
\end{equation}
where 
\begin{equation}
g=\frac{a m^{2}}{(m-1) Q_{Ni}}
\end{equation}

This is simply an approximation to the performance of supermirror data from neutron guide manufacturers' websites (e.g. \mbox{http://}\hspace{0 cm}\mbox{www.}\hspace{0 cm}\mbox{swissneutronics.}\hspace{0 cm}\mbox{ch}\hspace{0 cm}\mbox{/products}\hspace{0 cm}\mbox{/coatings.}\hspace{0 cm}\mbox{html}), for which $R_{Ni}\sim 0.98$ and modern coatings have $a\sim 0.01228$ giving $R_{crit} \sim 0.98 - 0.01228m^2$ as a good approximation for the reflectivity at the critical edge.

The Monte-Carlo method differs because it is possible to add more realistic details to the curve or even use a measured supermirror reflectivity profile.  In this case, VITESS was used with more realistic, rounded reflectivity curves at $Q_{Ni}$ instead of the piecewise function that equation \ref{eq:NADSreflectivity} generates.

For a fair comparison, in this study we have restricted the maximum width in the middle of the guide to that of the parabolically tapered ballistic guide as studied by Schanzer \emph{et al}, i.e. 0.36 m, with the same total length of 50 m.  We have modelled an 18.6 cm wide source similar to the Institut Laue-Langevin (ILL) horizontal cold source.  The effects of coma are largely wavelength independent neglecting critical angles for reflection, but for illustration purposes we simulate three Maxwellian curves of characteristic temperatures of 163 K, 382 K and 37 K, and brightnesses $1.67\times 10^{13}$, $3.97\times 10^{12}$ and $1.21\times 10^{13}$, again matching the ILL horizontal cold source, and a relatively monochromatic beam at 4 \AA.  Our aim is to maximise the neutron beam current striking a sample of area 4 cm $\times$ 4 cm without sacrificing the homogeneity of the phase space.  As we are interested in relative changes, we ignore the vertical plane and concentrate on the effects of varying the geometry in the horizontal plane.

This sample size is at the larger end of the sample size range for neutron instruments, but beam homogeneity over such a breadth is an important design feature of spectrometers such as LET at ISIS \cite{BEWLEY_LET}, for example.  It should further be noted that sample size does not affect the guide geometry and coma.  In the worst case, masking down a beam suffering from large coma causes a larger instrumental background and reduces the on-sample beam current compared to the hybrid guide.  In the worst case, it will still produce a tri-modal divergence distribution in the instrument resolution function.

\section{Elliptic Guides}
The literature on elliptic guides points back to an excellent article by Schanzer \emph{et al} \cite{SCHANZER_BALLISTIC}.  Schanzer compares a linearly tapered guide with a guide of the same geometry featuring parabolic tapering, and with a fully elliptical system.  In the present study, we deal only with the parabolic version of the ballistic guide (henceforth referred to as ``the ballistic guide'') and the ellipse.

\begin{figure}
\includegraphics[width=70mm]{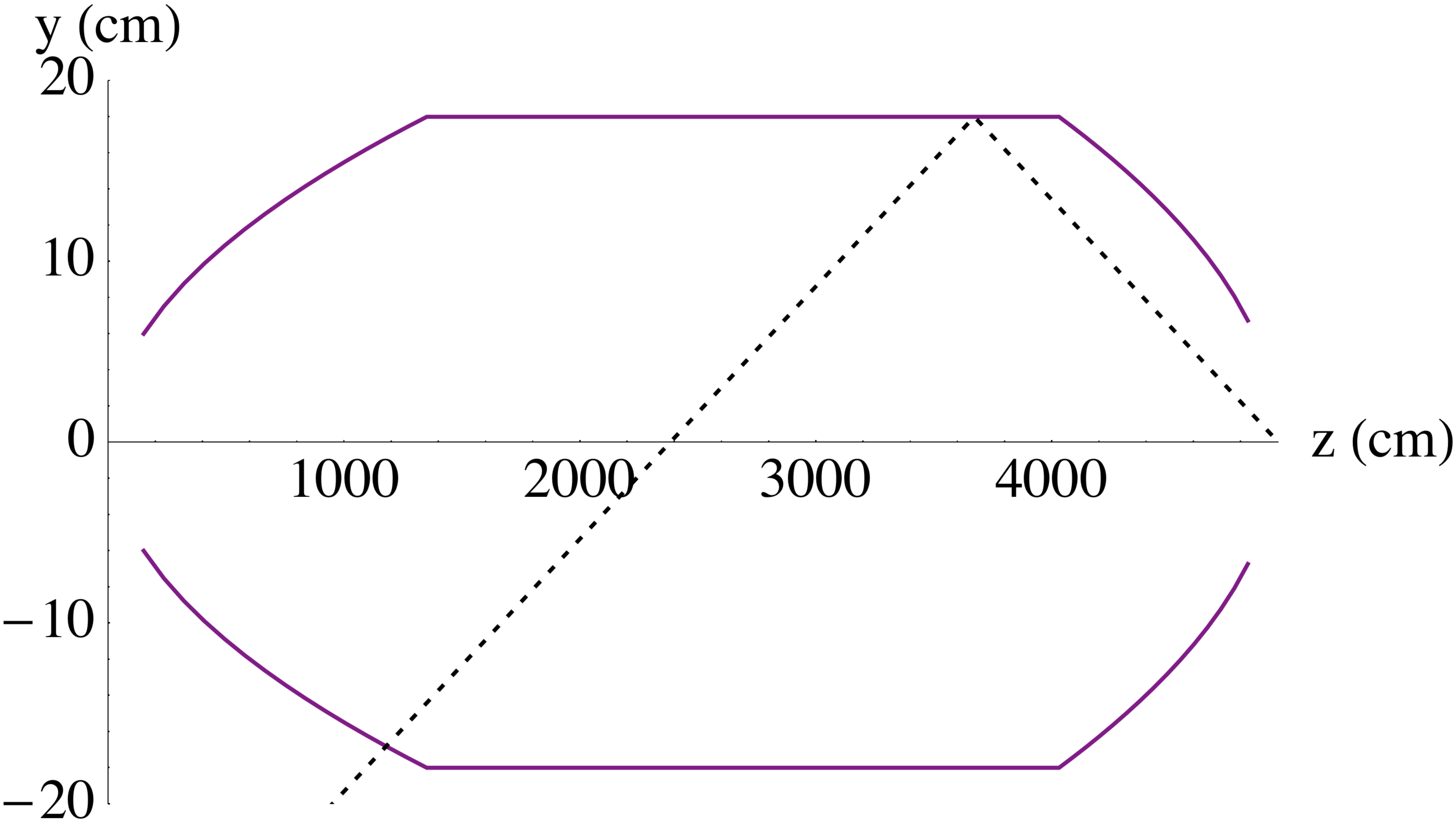}
\caption{Profile of the original ballistic guide for comparison with an ellipse, by Schanzer \emph{et al} \cite{SCHANZER_BALLISTIC}.  The broken line is a reverse-traced trajectory corresponding to one of the two minima in the divergence profile at $\sim 0.9^\circ$ shown in figure \ref{fig:ballisticDiv}.}
\label{fig:ballisticProfile}
\end{figure}

\begin{figure}
\includegraphics[width=70mm]{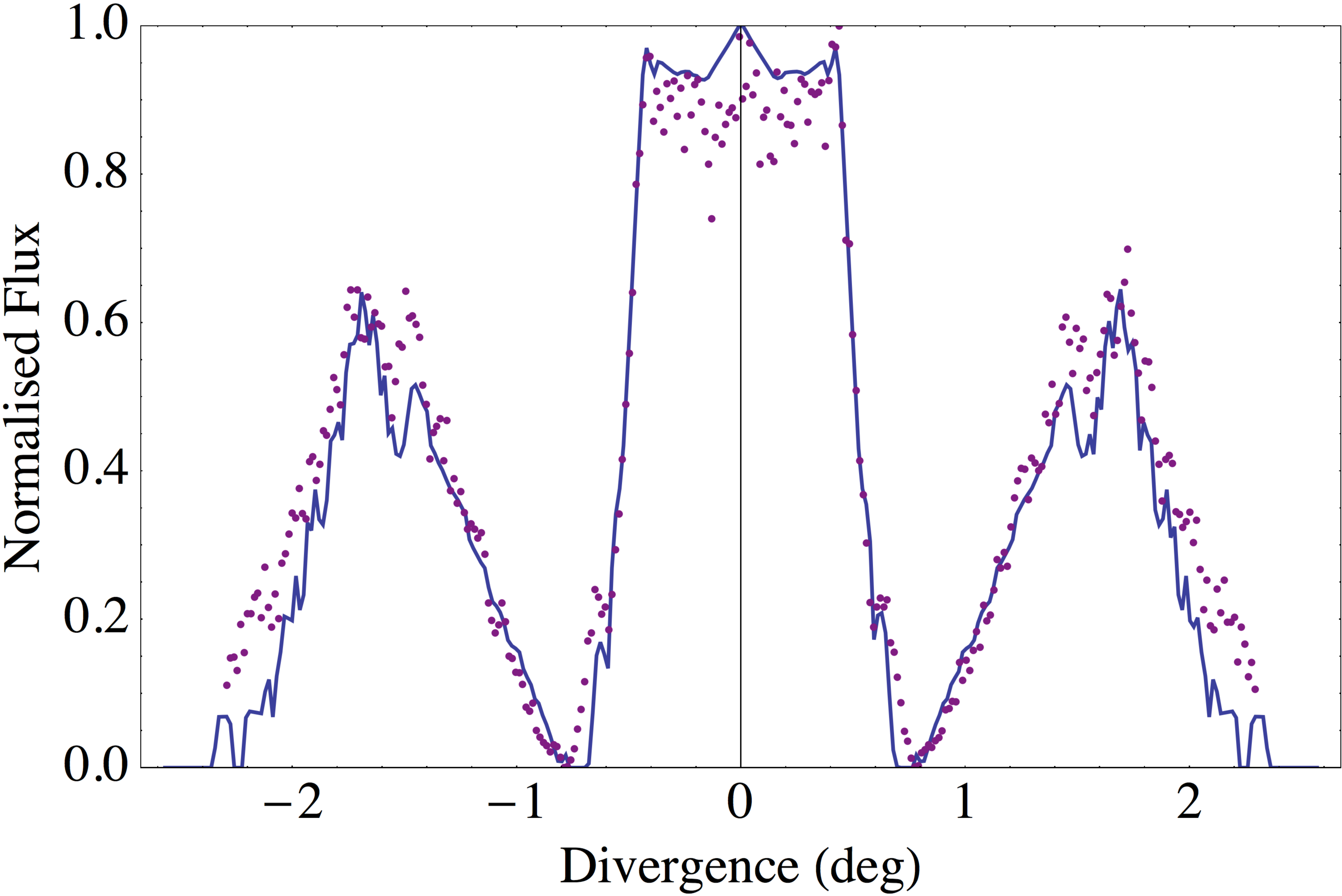}
\caption{The distribution of the beam divergence produced 1.3 m downstream of the exit of a ballistic guide like the one shown in figure \ref{fig:ballisticProfile}.  The solid lines are calculated using NADS \cite{BENTLEY_NADS} and the data points are computed using VITESS \cite{VITESS}.  Both curves are normalised so that the beam at zero divergence (i.e. direct view of the source) has a relative flux of one.  The maxima and minima in the NADS data are caused by the use of short, straight sections of guide to approximate curved surfaces, and exaggerated by the idealised reflectivity curve.}
\label{fig:ballisticDiv}
\end{figure}

Figure \ref{fig:ballisticProfile} shows a typical ballistic guide profile.
This geometry produces an inhomogeneous divergence distribution, shown in figure \ref{fig:ballisticDiv}.
When tracing backwards from the target, it becomes clear why there are two holes in the phase space.  The expanding parabola nearest to the source would have to reflect at large angles to supply this trajectory with a neutron.

Such a trimodal divergence profile has been observed before \cite{GONCHARENKO_BALLISTIC}.  With three independent beams crossing at the sample, the angular component of the resolution function in one instrument plane might require at least six parameters --- the angle of incidence and width of each beam component.  Beyond the scope of this discussion, but worth bearing in mind, are the implications for chopper transmission functions in time-of-flight spectrometers, because of the correlations between divergence and position in the guide phase space.

Note that simply increasing the $m$ value does not completely solve the problem, because at high-$m$ and high angles the reflectivity is significantly lower than unity.  Although the holes in phase space may be reduced in depth, they still remain.  Ultimately, a more homogeneous beam requires a change in the geometry of the entire guide system.

\begin{figure}
\includegraphics[width=70mm]{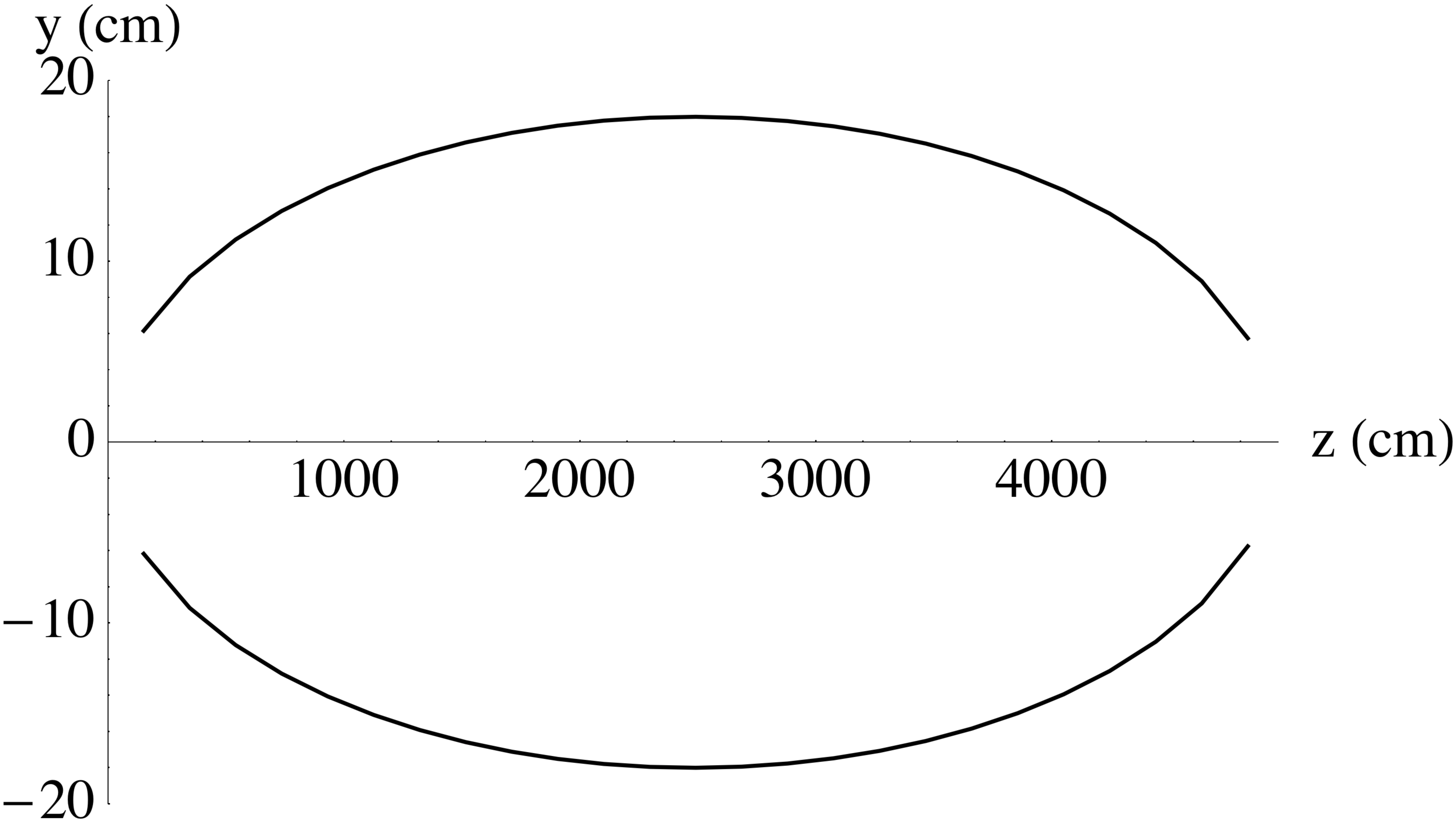}
\caption{Profile of an elliptic guide similar to that described by Schanzer \emph{et al} \cite{SCHANZER_BALLISTIC}.}
\label{fig:ellipseProfile}
\end{figure}

\begin{figure}
\includegraphics[width=70mm]{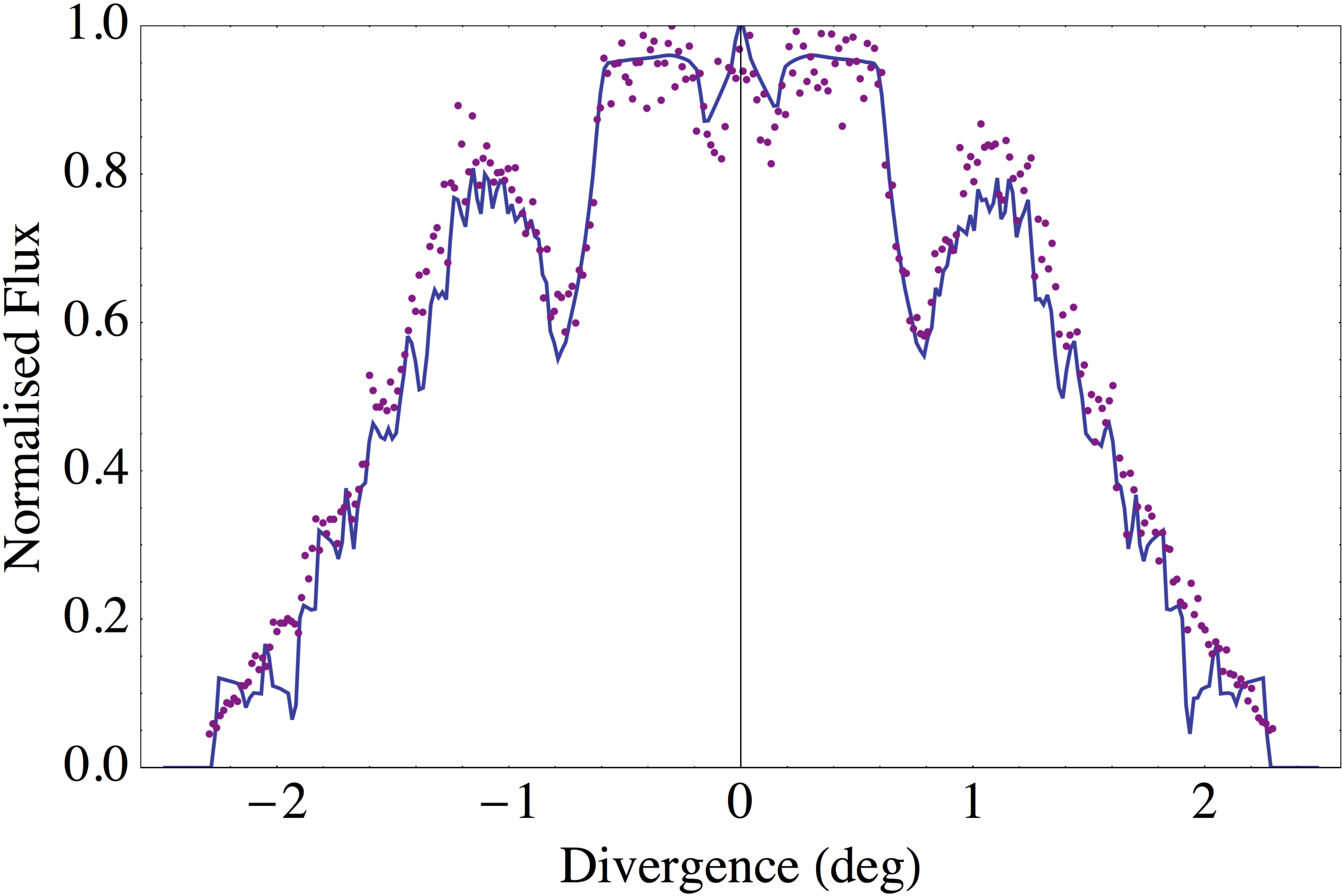}
\caption{The distribution of the beam divergence at the target position as produced by an elliptic guide.  The solid lines are calculated using NADS and the data points are computed using VITESS.  Both curves are normalised so that the beam at zero divergence (i.e. direct view of the source) has a relative flux of one.  The maxima and minima in the NADS data are caused by the use of short, straight sections of guide to approximate curved surfaces, and exaggerated by the idealised reflectivity curve.}
\label{fig:ellipseDiv}
\end{figure}

The solution found by Schanzer is to exploit the geometry of an ellipse, shown in figure \ref{fig:ellipseProfile}.
As mentioned previously, an elliptic mirror has the property in that any ray emitted from a point source at one focal point is reflected once and only once, and arrives precisely focused at the other focal point (neglecting gravity).  Figure \ref{fig:ellipseDiv} shows that the beam divergence distribution is much smoother in an elliptic guide compared to that of the ballistic guide in figure \ref{fig:ballisticDiv}, even with a basic design featuring a uniform $m$ value throughout the length of the guide.
This smoothness or uniformity of the divergence distribution is what we refer to as ``beam quality''.  In addition to the improvement in beam quality, the elliptic system provides a large increase in beam transport efficiency relative to the straight or ballistic system.

This improvement in beam quality can be expected \emph{for some guide geometries}, but it would be a mistake to treat this result as a general case.  We stress this point because firstly the parabolic-ballistic system in Schanzer's paper is not optimal.  Secondly, linearly tapered guides can also be designed to reduce the problems shown in figure \ref{fig:ballisticDiv}, and thirdly it is easy to misconfigure an ellipse such that it also produces very poor quality beams.

\begin{figure}
\includegraphics[width=70mm]{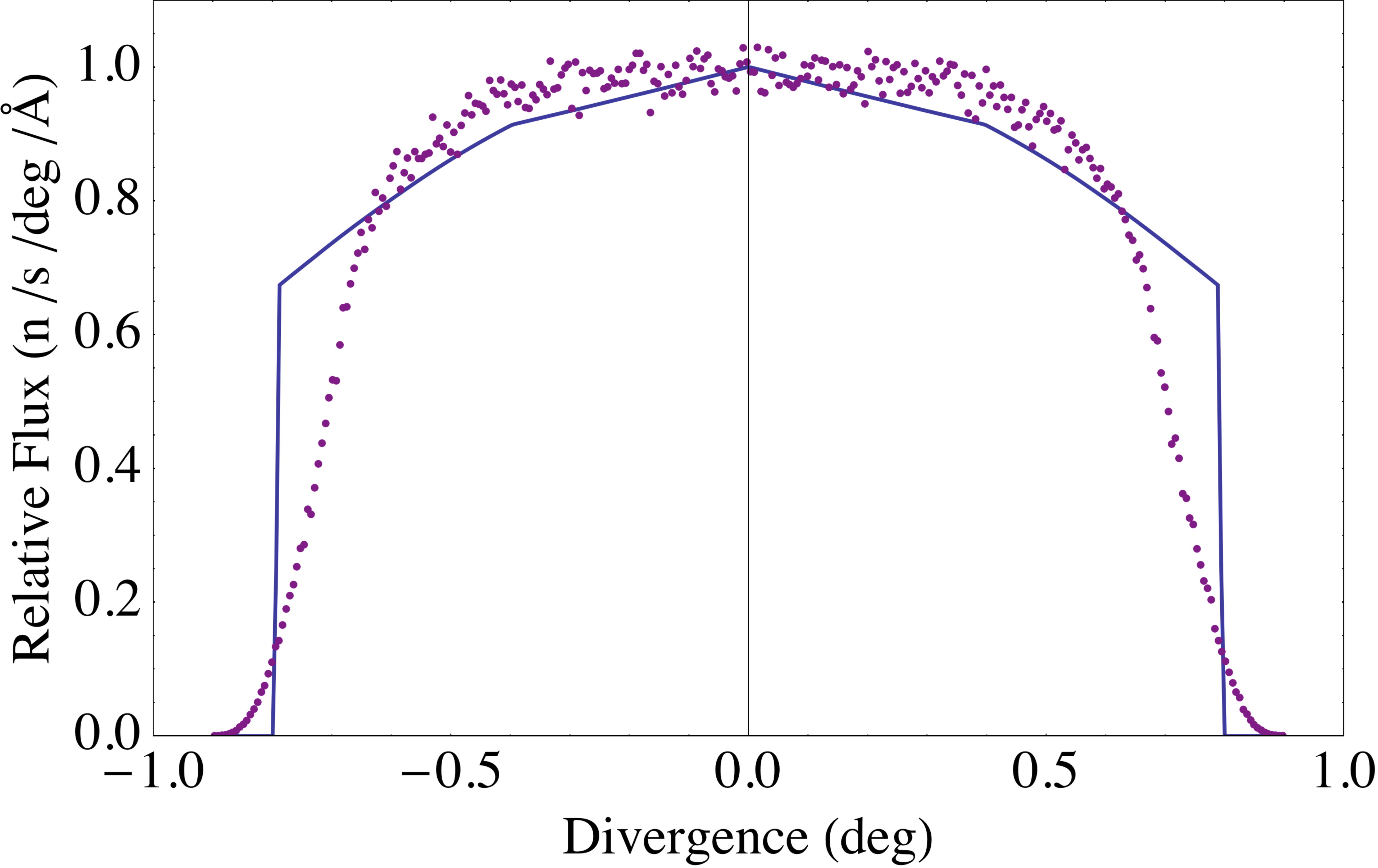}
\caption{Beam divergence distribution of a straight neutron guide.  The solid lines are calculated using NADS and the data points are computed using VITESS.  Both curves are normalised so that the beam at zero divergence (i.e. direct view of the source) has a relative flux of one.}
\label{fig:straightDiv}
\end{figure}

The elliptic divergence distribution shown here is a great improvement over the ballistic guide.  Figure \ref{fig:straightDiv} shows that it is still significantly worse than the beam quality produced by a straight guide.
This is important, because resolution calculations in neutron scattering often make idealised assumptions about simple nature of the divergence distribution of the beam (see, for example, the article by Loong \emph{et al} \cite{LOONG_CHOPPER_RESOLUTION}).  However, it is possible to design an elliptic guide with greatly enhanced intensity and with little loss in divergence compared to the straight guide \cite{BOENI_ELLIPTIC_GUIDE_SHIELDING}.

\section{Coma in Elliptic Guides}

One intrinsic problem with an elliptic neutron guide, and indeed any curved mirror, is the issue of coma.  Coma is a well-understood phenomenon in reflecting optics.  Early designs of reflecting telescopes, and cheap modern ones, produce images of star fields with sharp star images in the centre, but stars at the edges of the image are not well resolved.  Instead, they appear to have tails resembling those of comets, hence the name ``coma''.  Coma becomes a serious issue for neutron instrument resolution and background when one attempts a full design study of a neutron scattering instrument, such as small-angle neutron scattering instrument (SANS) \cite{RODRIGUEZ_ELLIPTIC_SANS}.

\begin{figure}
\includegraphics[width=75mm]{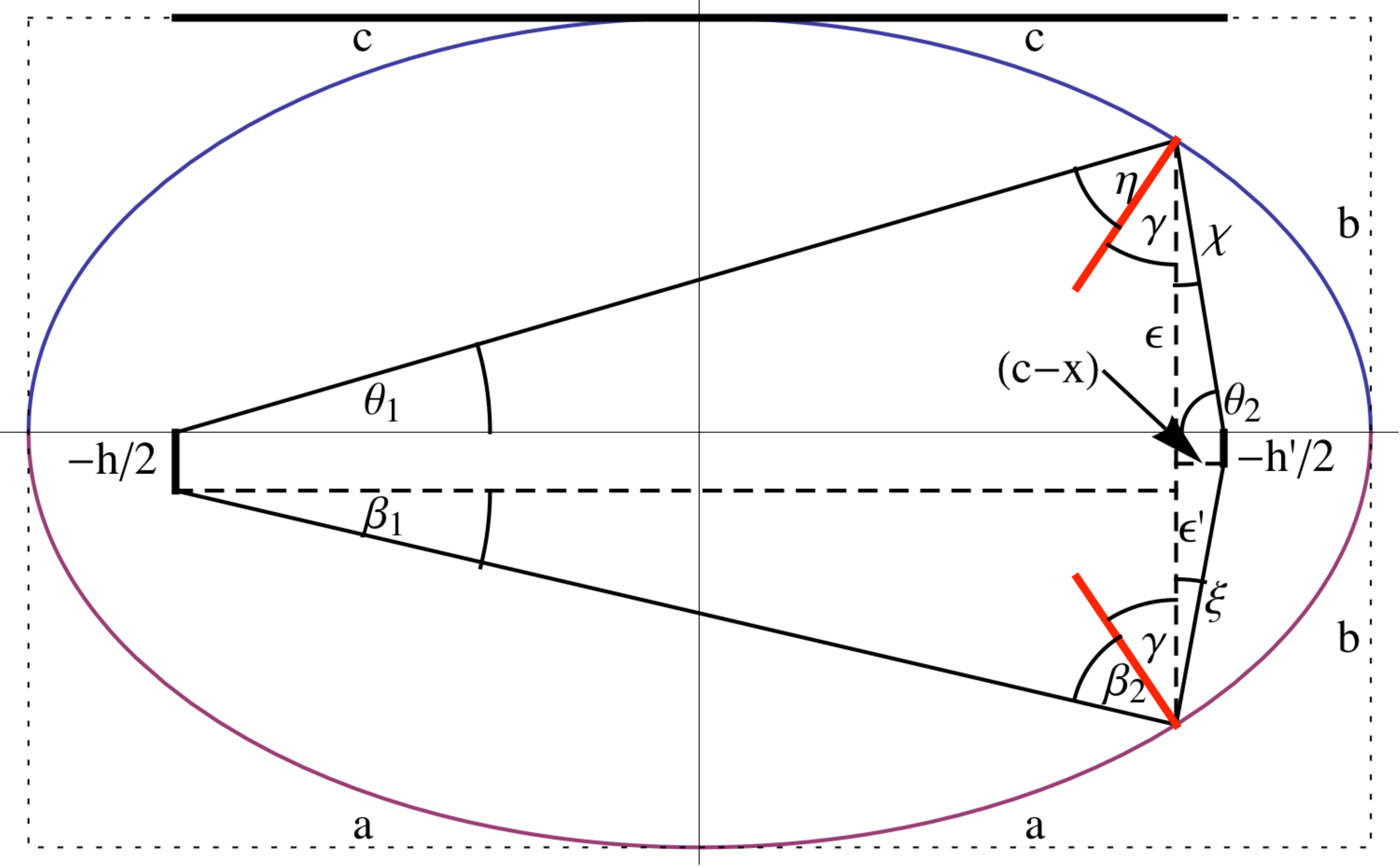}
\caption{The origin of coma in elliptic guides.  The trajectory in the upper half plane of the figure shows a ray from an on-axis focal point source correctly focuses onto the opposite focal point.  The lower half plane shows that a similar ray from an off-axis point on the source that is reflected at the same horizontal position $x$ is not brought to the same focal point (where $x=0$ is at the ellipse centre).  Rays reflected in the further half of the ellipse create a smaller blurred image of the source, whereas if rays were moving in the reverse direction reflecting in the near half of the ellipse they would create a large blurred image.  The frame of the figure also shows the semimajor and semiminor axes $a$ and $b$, and the half-distance between the foci, $c$.}
\label{fig:ellipseComa}
\end{figure}

Here, we do not derive coma strictly, but illustrate how the problem arises using figure \ref{fig:ellipseComa}.
This compares two trajectories very simply, one from the leftmost focal point of an ellipse, and one from a point directly below at a distance $h/2$ to simulate the effect of a source with height $h$.  Navigating around figure \ref{fig:ellipseComa} allows the derivation of the spatial extent of the image as a function of the location of the reflection in the horizontal direction $x$, which is valid in the small grazing angle regime for neutron guides: 

\begin{eqnarray}
h' =   &2&  \left\{\epsilon+(c-x) \tan \left[\tan ^{-1}\left(\frac{h-2
   \epsilon }{2 (c+x)} \right) \right. \right.  \nonumber \\
     &+{}& 2 \tan^{-1}\left(\frac{c-x}{\epsilon }\right) \nonumber \\
   & +{} & \tan^{-1}\left(\frac{\epsilon }{c-x}\right) \nonumber \\
   & + {} & \left. \left. \tan^{-1}\left(\frac{\epsilon
   }{c+x}\right)\right] \right\}
   \label{eqn:comaDerivation}
\end{eqnarray}
where
\begin{equation}
\epsilon = \sqrt{b^2-\frac{b^2 x^2}{b^2+c^2}}
\end{equation}

and where $a$ and $b$ are the semimajor and semiminor axes respectively, and $c$ is half the distance between the two foci (i.e. the distance from the ellipse center to a focus) given by $c^2 = a^2 - b^2$.

Equation \ref{eqn:comaDerivation} is very well approximated \cite{RODRIGUEZ_ELLIPTIC_SANS} by
\begin{eqnarray}
\label{eqn:simpleComa}
h' & \approx  & h\cdot (c-x)/(c+x) \nonumber \\
 & \approx & h\cdot (a-x)/(a+x)
 \end{eqnarray}
 because we are in the small angle regime and $a\approx c$ and $b \ll a, c$.

\begin{figure}
\includegraphics[width=70mm]{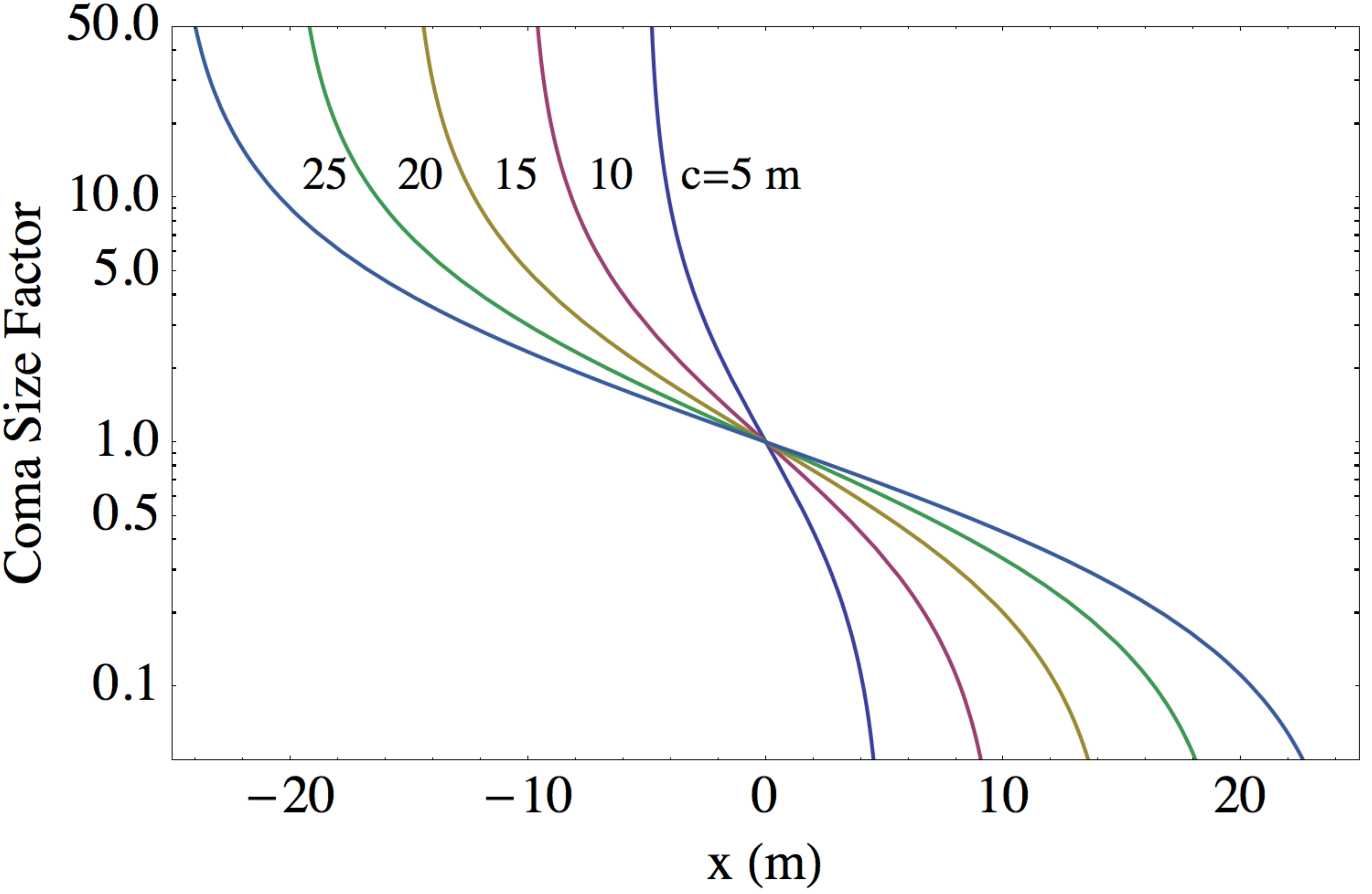}
\caption{Spatial extent of the coma blurring $(h'/h)$ as a function of reflection position $x$.  Several values of $c$ are given, which is half the distance of between the focal points in metres.  Here we see that the worst coma effect is caused by rays striking the elliptic mirror close to the extended source, and after the mid-way point the coma effect is inverted so that the image is smaller than the source.  For this figure, the semiminor axis $b$ is fixed at 0.18 m, so that the guide length is determined by $a^2 = c^2 + b^2$, although the $b$-dependence of $(h'/h)$ is essentially zero in this geometry regime (see equation \ref{eqn:simpleComa} and surrounding text).}
\label{fig:comaMag}
\end{figure}

If an elliptic mirror were not to suffer from coma, then $h'/h$ would be independent of $x$.  Figure \ref{fig:comaMag} shows the $x$-dependence of $h'/h$, which can be interpreted as the spatial extent of the blurring caused by coma --- a ``coma size factor''.  The figure shows a range of distances between the focal points, revealing that the size of the blurred image for any given neutron trajectory depends upon where exactly along the guide the neutron is reflected in relation to the entrance, the mid-point and the exit, but not on the length of the guide ($2a$) or the distance between the focal points ($2c$) in absolute terms.  In these graphs, we have used $b=0.18$ m, i.e. the ellipse is 36 cm at the widest point, but in this geometry regime $a\gg b$, the coma blurring is essentially independent of $b$ also.

With the source focus on the left and the sample focus on the right of figure \ref{fig:ellipseComa}, then figure \ref{fig:comaMag} shows that for the rays reflected in the guide sections nearest to the source, a 1 cm wide sample would be expected to produce a blurred image with tails that are 20-50 cm wide.

These are reasonably well-collimated rays that strike the ellipse a second time further down the guide system irrespective of the choice in focal points.  This multiple reflection phenomenon from the coma is what causes the features at high angles in figure \ref{fig:ellipseDiv}, and we refer back to item \ref{cause:reflectivityCurve} in the list of design problems in the introduction section.  The central maximum region is from the correctly focused rays from the source, the first half of the ellipse, and the central region of the ellipse.  The minima at $\pm 0.9^\circ$ correspond to high grazing angles for the reflection closest to the target.  The lobes at $\pm 1.2^\circ$ are caused by multiply reflected coma trajectories at relatively low grazing angles for both reflections, and finally the long tails are from the extreme grazing angles of the incoming rays.  These features are of course amplified a little by our intentional use of only one $m$ value throughout the system, but the data in figure \ref{fig:ellipseDiv} are not inconsistent with those in Schanzer's study.

One possible solution to the coma problem is to make a pseudo point source with an absorbing beam mask.  This is not an ideal solution because the tails of the coma at the sample position still extend to many times the size of the mask aperture.  More importantly, by masking the source the incident flux is considerably reduced from its potential level by perhaps more than an order of magnitude, negating any flux benefits of using an elliptic guide over a straight guide.

For simplicity, we have ignored several other effects that contribute to the blurring of the image.  Guide waviness is a minor perturbation that, for 50 m long guides, is expected to contribute a blurring of a few 10's of mm or smaller --- much smaller than the coma effect --- but it does not change the underlying guide shape and therefore does not remove coma from the ellipse.  It should be noted that varying $m$, or improving the polygonal approximation to the ellipse with many more straight sections, or even continuous mirrors, also do not change the underlying shape of the elliptic mirror and therefore do not remove the coma effect.  We seek a more general solution to coma that can be applied to any elliptic guide deployment if required.

\section{Eliminating Elliptic Guide Aberrations}
A far better solution would be to eliminate or reduce the effects caused by the aberration.  In telescopes, this is achieved by carefully designing a secondary mirror to reduce or eliminate any optical aberrations caused by the primary mirror.  Wolter \cite{WOLTER_OPTICS} has designed a number of low grazing angle optical device types which serve precisely this purpose.  They have been demonstrated to work excellently for both x-rays \cite{CHANDRA} and neutrons \cite{GUBAREV_WOLTER,MILDNER_WOLTER,KHAYKOVICH_WOLTER}.

\begin{figure}
\includegraphics[width=70mm]{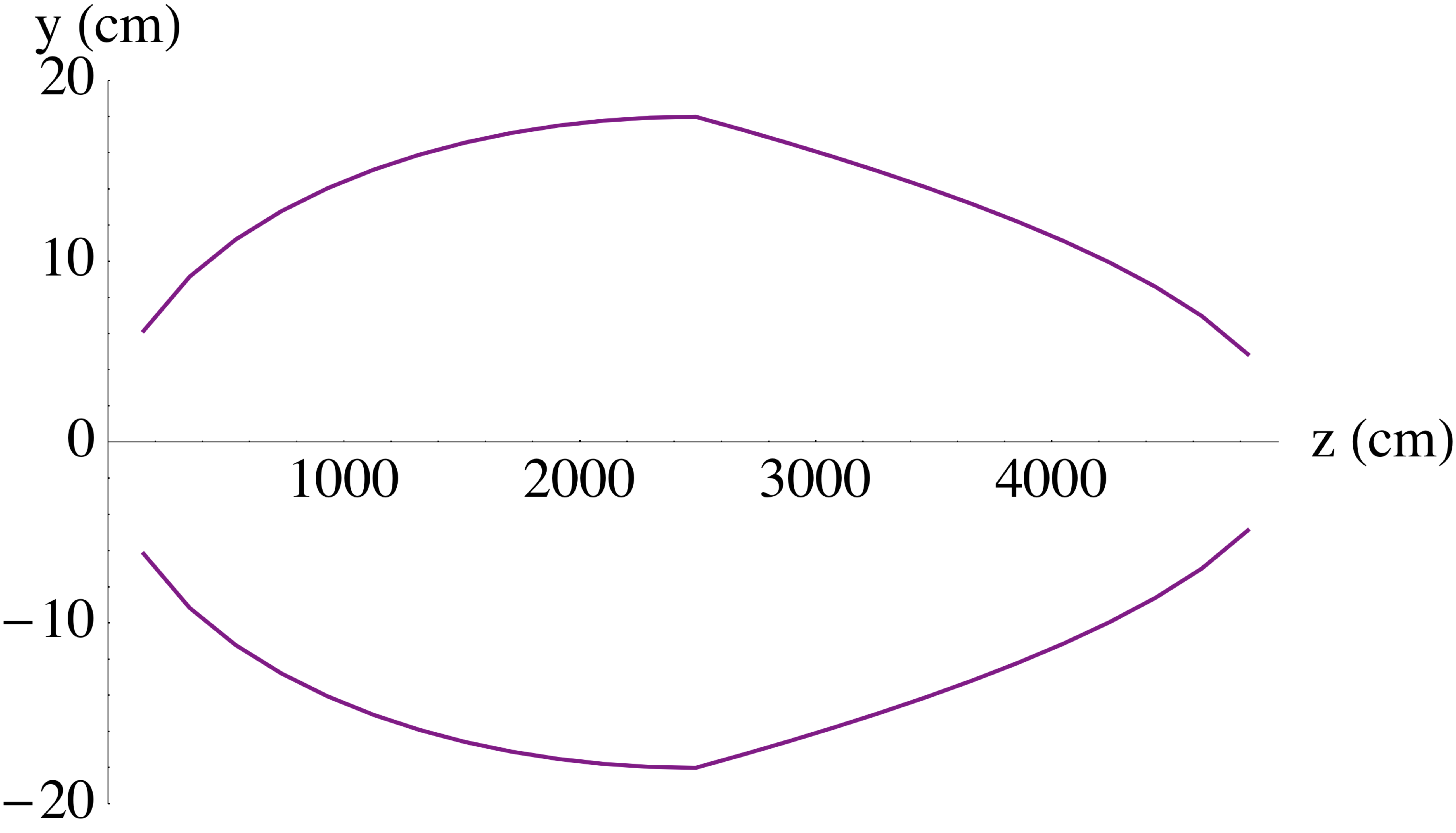}
\caption{Profile of a hybrid guide offering maximal quality in beam divergence, for a penalty of 10-20\% in neutron beam current.}
\label{fig:hybridQualityProfile}
\end{figure}

Our requirements in this study are not quite as strict.  We are not necessarily interested in obtaining a point image of the beam, but improvements over elliptic guide systems to remove the multi-modality in the divergence distributions without compromising the beam transport performance of the system.  To this end, we have experimented with several configurations of hybrid guide system, where the first half of the guide has one particular conic section type, and the second part of the guide uses a different conic section type.  A survey of the 6 possible permutations (elliptic-parabolic, elliptic-hyperbolic, parabolic-elliptic, parabolic-hyperbolic, hyperbolic-elliptic, hyperbolic-parabolic) reveals that the elliptic-parabolic hybrid system offers the best performance for this particular case.  This should not be treated as a general result --- it is possible that a different combination of conic section offers better performance depending on the distances involved, and the spatial extent of the source and target.  In any case, we will now focus our attention on this optimal elliptic-parabolic configuration (henceforth named ``hybrid guide'' for simplicity) and compare it with the current guide geometries.  Figure \ref{fig:hybridQualityProfile} illustrates this hybrid guide concept with an example that has the transition from ellipse to parabola at the mid point of the guide system.

To fully optimise a hybrid guide, there are more degrees of freedom than a regular elliptic guide:
\begin{enumerate}
\item The first focal point of the elliptic section
\item The second focal point of the elliptic section
\item The location of the transition from elliptic to parabolic shape
\item The focal point of the parabola
\item The maximum width of the guide
\end{enumerate}

The placement of the focal points do not necessarily have to coincide with the source and target, and in fact it is frequently optimal in neutron guides to have the focal points farther from the guide entrance/exit planes than the source and target.

The crossover point gives a degree of control on the homogeneity of the divergence profile from the guide system.  In the limit of crossover at the guide exit, it is a purely elliptic system, and with the crossover at the guide entance it is purely parabolic.  Crossover in the exact middle of the guide, as shown in figure \ref{fig:hybridQualityProfile} provides a very high beam quality in terms of homogeneity in the divergence distribution.  We found in this case that this incurred a small cost of around 10-20\% in total flux relative to a pure ellipse.

\begin{figure}
\includegraphics[width=70mm]{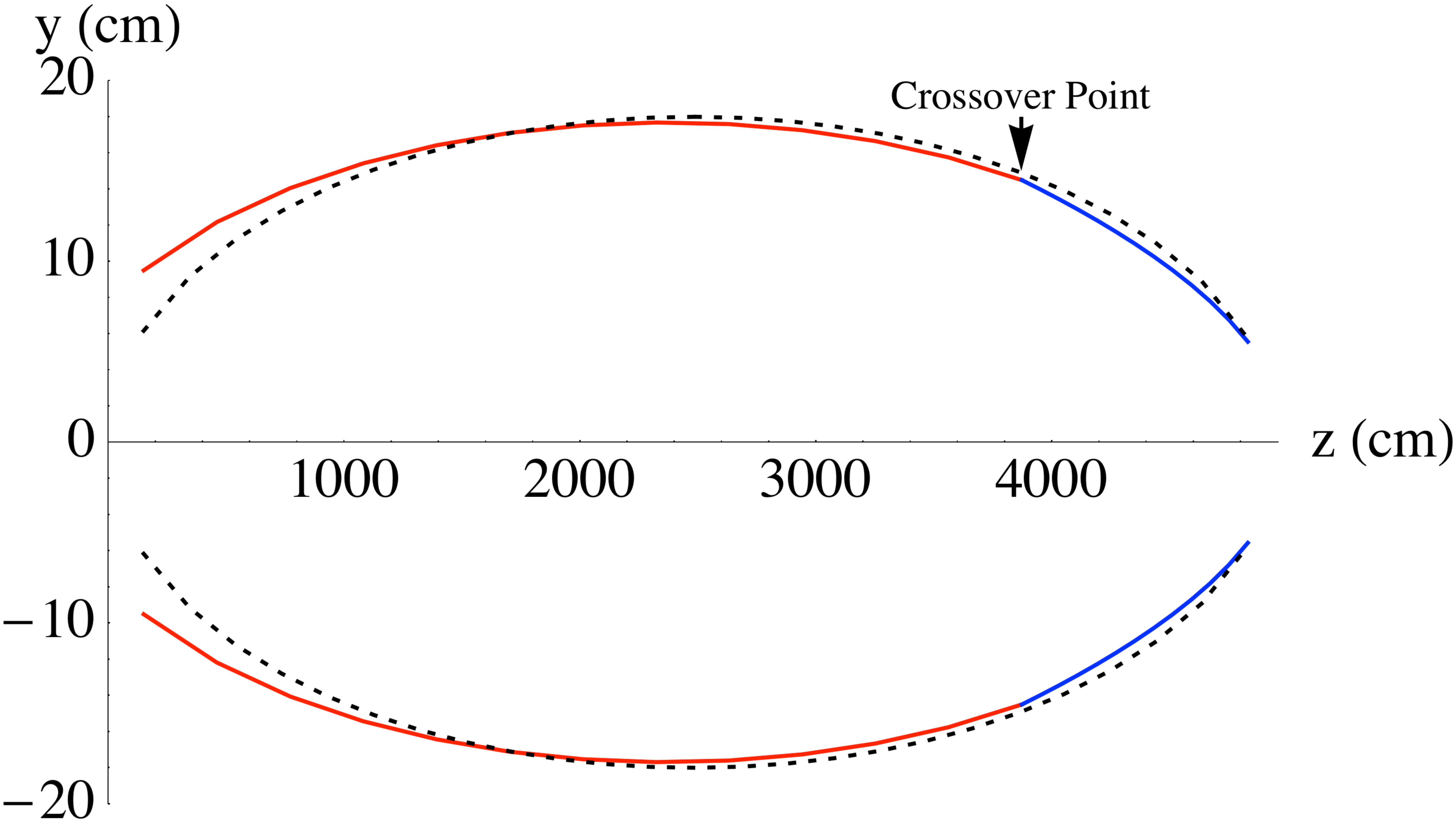}
\caption{Profile of a fully optimised elliptic-parabolic hybrid guide (solid lines), compared to that of a regular elliptic guide (dotted lines).  The cross-over point between the two types of conic section is around three quarters of the way along the length of the guide, in this case, marked with a change of colour.}
\label{fig:hybridProfile}
\end{figure}

The ``sweet spot'' for this hybrid geometry in the current project is found with the crossover at 80\% of the way between the entrance and exit.  This configuration was found by first optimising the geometry purely for beam current using a particle swarm (exactly as used previously in focusing neutron optics \cite{BENTLEY_MULTIBLADE}), and then manually adjusting the crossover to increase the beam quality until the beam current begins to decrease.  This provides maximum performance and good homogeneity in beam divergence.  Our final optimised geometry is shown in figure \ref{fig:hybridProfile}, where it is also compared with a pure elliptic geometry.  We see that the most notable deviation from the baseline ellipse is at the entrance, with a focal point significantly behind the source.

\begin{figure}
\includegraphics[width=70mm]{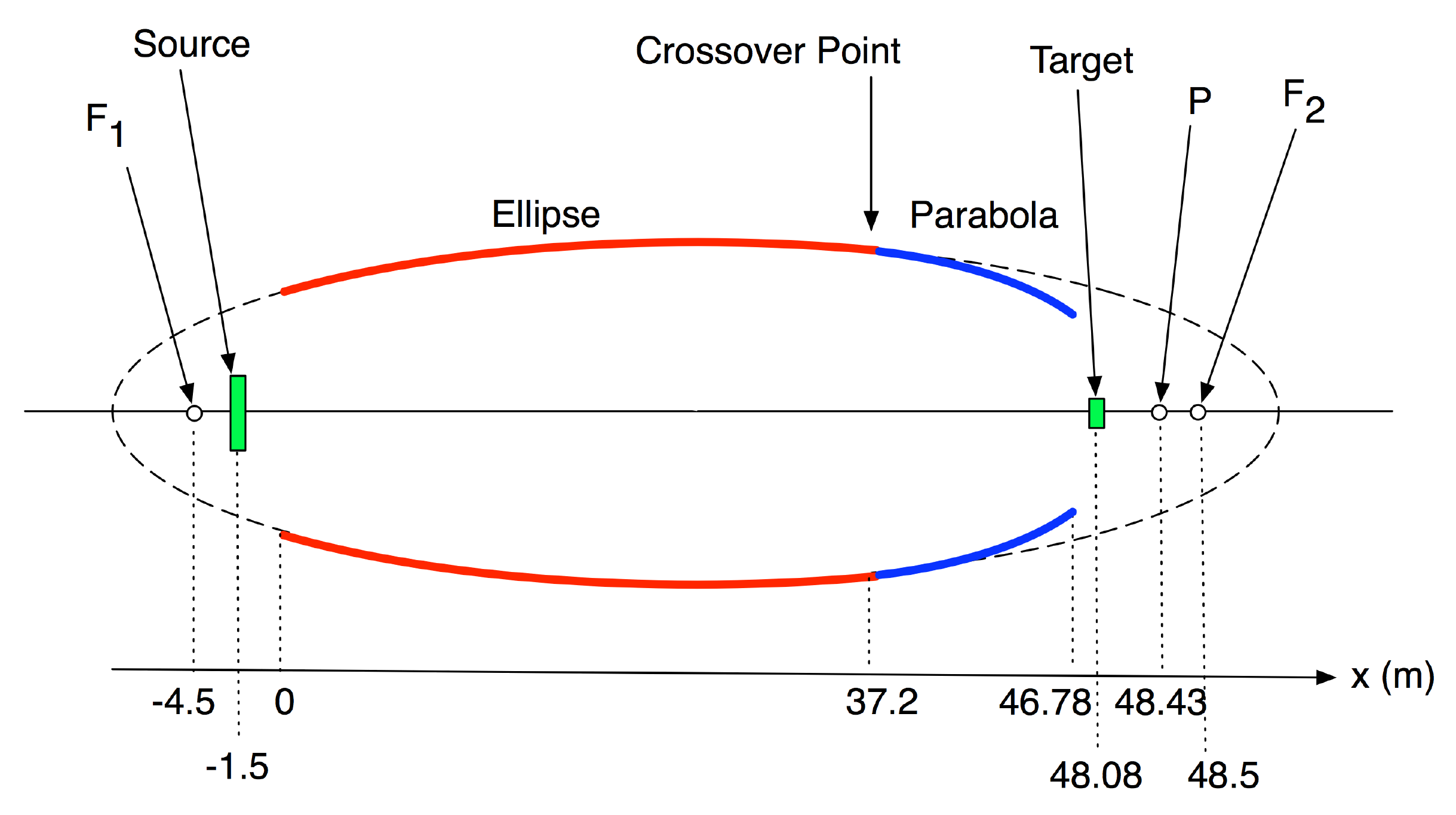}
\caption{Schematic of a hybrid guide illustrating the values of the parameters in table \ref{tab:hybridParams}.  Distances along the length of the guide are given in metres.  The elliptic part is on the left (red in colour); the parabolic part is on the right (blue in colour); and the dotted line is a full ellipse that overlaps with the geometry of the elliptic section.}
\label{fig:labelledHybridSchematic}
\end{figure}

The full description of the hybrid guide parameters is given in table \ref{tab:hybridParams}.  Figure \ref{fig:labelledHybridSchematic} is a schematic diagram showing these parameters on the optimised hybrid system.

\begin{table}
\caption{\label{tab:hybridParams} Parameters found for the optimised hybrid guide.  Focal points are labelled in parenthesis with the same symbols as those used in figure \ref{fig:labelledHybridSchematic}.}

\begin{tabular}{ll}

Parameter &  Value \\

\hline
Elliptic focal point 1 $(F_1)$ & -4.12 m \\
Source position & -1.5 m\\
Entrance to Elliptic Guide & 0 m \\
Elliptic focal point 2  $(F_2)$ & 48.5 m \\
Crossover point & 37.2 m \\
Parabolic guide exit & 46.78 m \\
Target & 48.08 m \\
Parabolic focal point $(P)$ & 48.43 m \\
Maximum width at widest point & 0.36 m \\

\end{tabular}
\end{table}

\begin{figure}
\includegraphics[width=70mm]{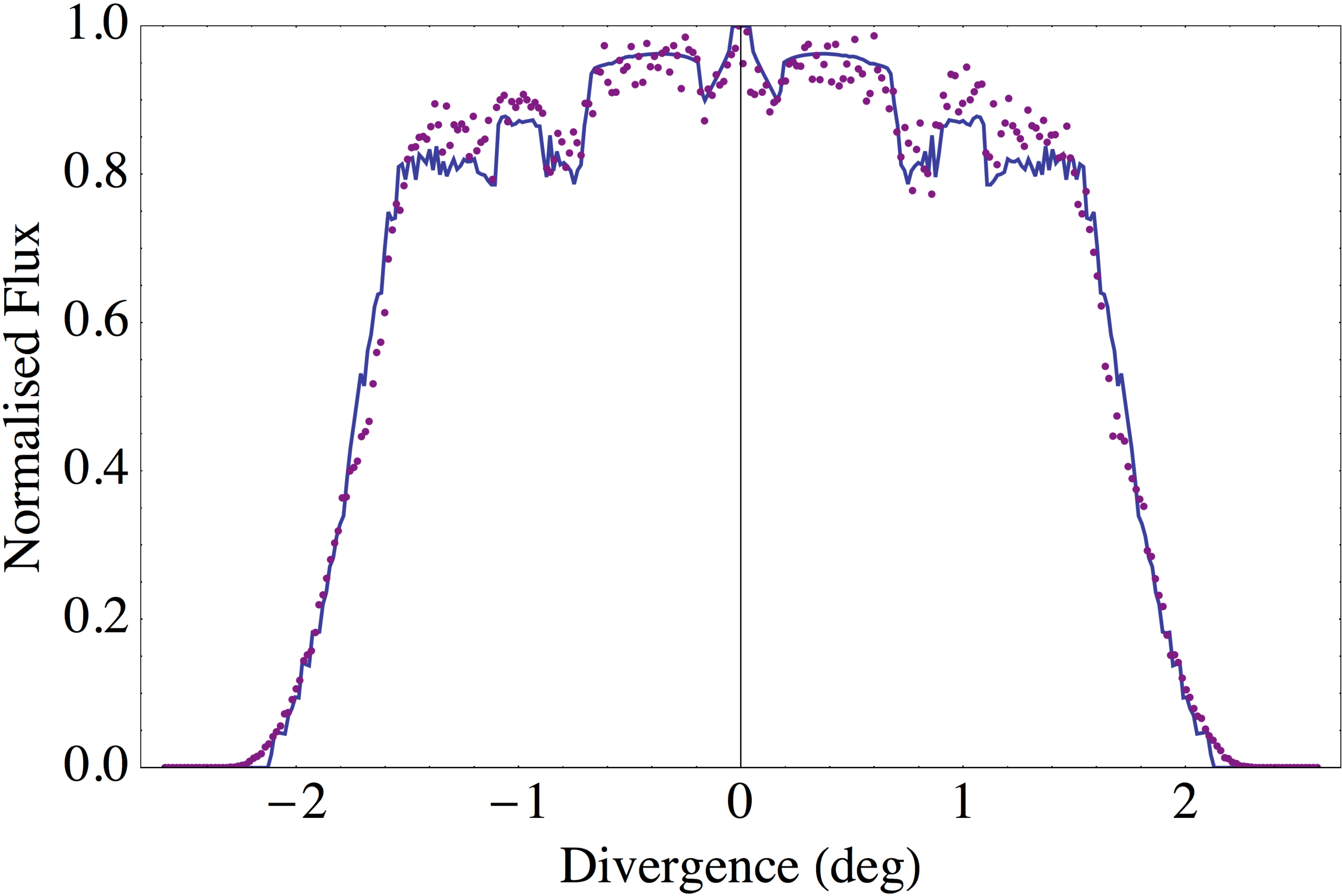}
\caption{Beam divergence distribution at the target position as produced by the hybrid guide.  The solid lines are calculated using NADS and the data points are computed using VITESS.  Both curves are normalised so that the beam at zero divergence (i.e. direct view of the source) has a relative flux of one.  The maxima and minima in the NADS data are caused by the use of short, straight sections of guide to approximate curved surfaces, and exaggerated by the idealised reflectivity curve.}
\label{fig:hybridDiv}
\end{figure}

The result of these slight changes are apparent in Figure \ref{fig:hybridDiv}, which shows that our hybrid system has a much smoother divergence distribution.
Indeed, the beam quality is comparable to that of a straight neutron guide, shown in figure \ref{fig:straightDiv}, although it matches the elliptic guide for both divergence and high beam transport.

\begin{figure}
\includegraphics[width=70mm]{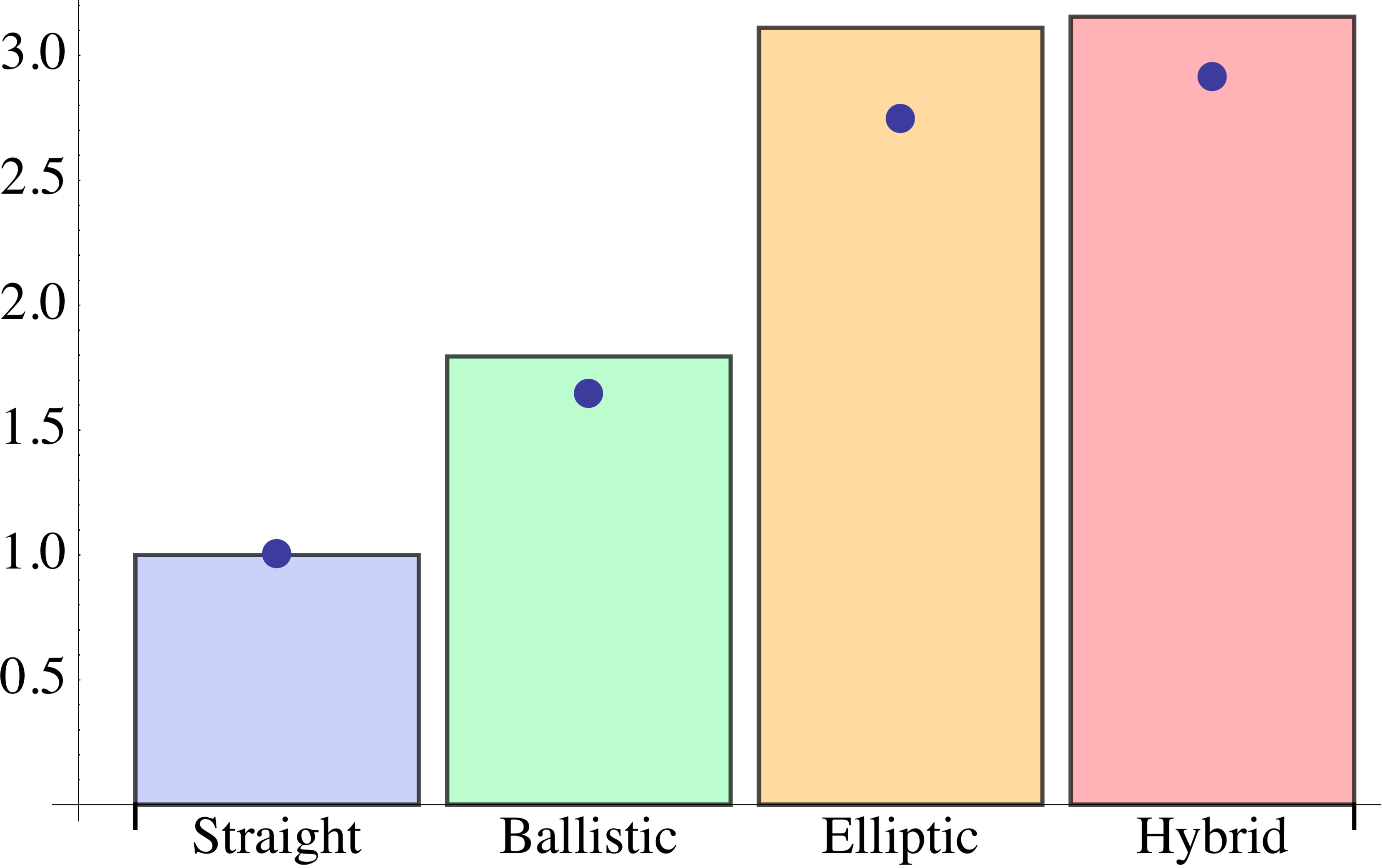}
\caption{On-sample calculated flux for each of the geometries relative to the straight guide, for the same length.  The bars are the performance as calculated using NADS, the points were calculated using VITESS.}
\label{fig:performance}
\end{figure}

Figure \ref{fig:performance} compares the flux performance of each of the guide geometries that have been modelled.  The gains of course come from an increased divergence of the beam, so a fair comparison must also take into consideration the useful divergence of a spectrometer coupled to the guide, bearing in mind figures \ref{fig:ellipseDiv}, \ref{fig:straightDiv} and \ref{fig:hybridDiv}.

Here we see that the ellipse and the hybrid have very similar flux gains relative to the simple, straight guide, and the hybrid geometry may even have a slight edge over the ellipse.

\section{Discussion}

Despite the fact that a hybrid guide involves two reflections, it can remove the effect of optical aberrations that are present in a full ellipse, i.e. a system built from a single conic section type.  This is not surprising when one considers the understanding of aberrations that is employed  in Wolter optics and modern reflecting telescope design.  What is surprising here, in the context of neutron guides, is that the best performance is achieved by using principally the \emph{worst} part of the ellipse for optical aberrations.  With a little further examination, however, this hybrid system becomes very logical.

Firstly, in the mid-section of the guide we see from figure \ref{fig:hybridProfile} that there is little difference between the ellipse and the hybrid system.  Here, the size of the coma effect is negligible.  The second notable feature from the optimisation of the hybrid guide is that the first focal point of the ellipse is a long way upstream of the guide entrance.  This reduces the grazing angle of the source rays that will suffer most from the coma, reducing reflectivity losses.

\begin{figure}
\includegraphics[width=70mm]{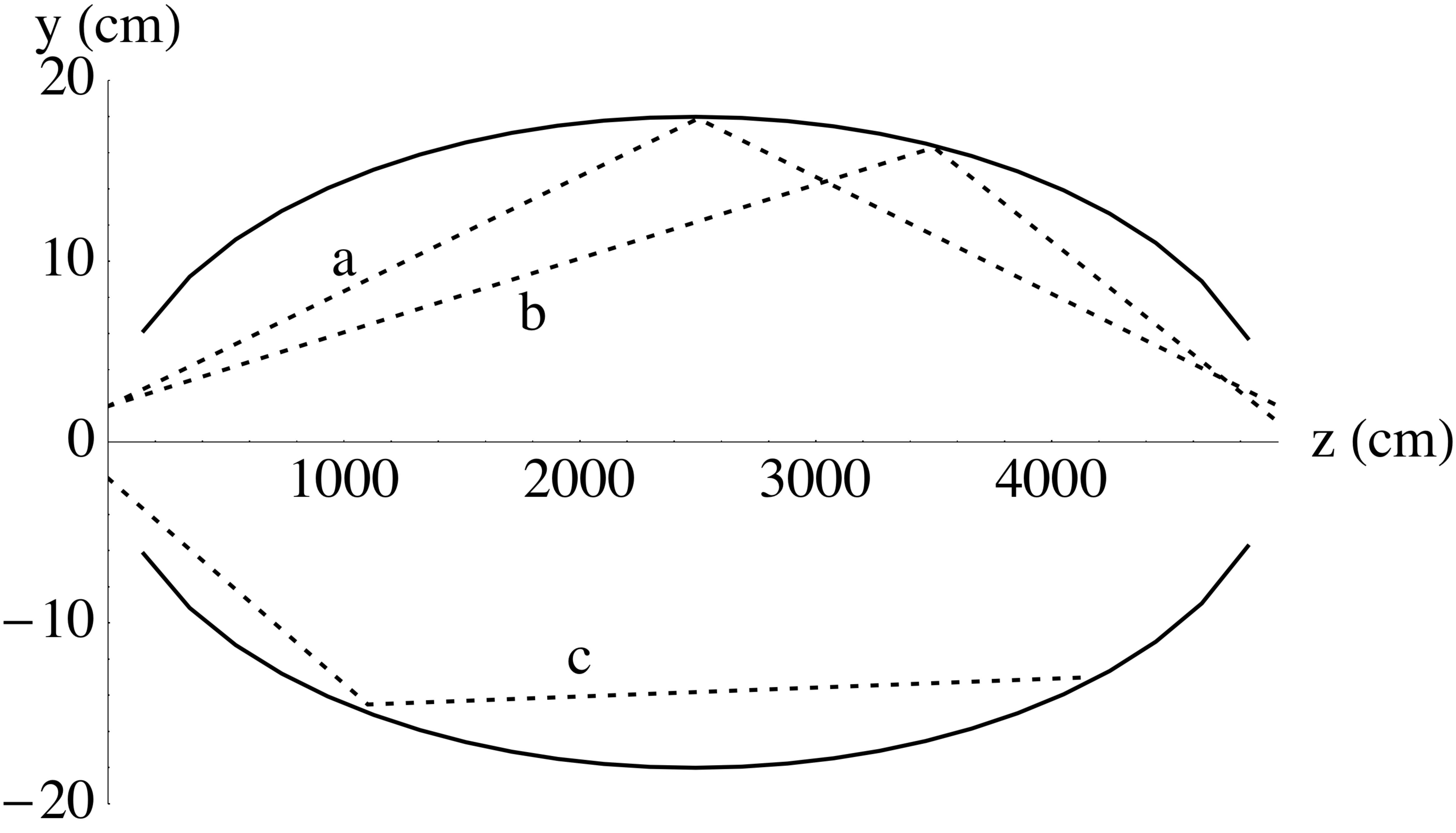}
\caption{Illustration of three off-axis trajectories in a regular elliptic guide.  ``a'' shows a reflection in the mid point, so that the coma size equals the source size.  ``b'' shows a reflection near the target, so that the coma size is smaller than the source size.  ``c'' shows a reflection near the source, producing multiple reflections in the guide.  These kinds of ``c'' trajectories are more efficiently handled by the parabolic part of the hybrid system.}
\label{fig:comaTrajectories}
\end{figure}

In both the full ellipse and the hybrid guide, these rays from the coma effect cause neutrons to undergo multiple reflections downstream, as shown in figure \ref{fig:comaTrajectories}.
In the elliptic case, the coma would not be corrected by these subsequent reflections, and an ellipse is most efficient at reflecting trajectories from the point source, not from the off-axis regions which are relatively well collimated.  In contrast, a hybrid system presents these rays with a parabolic section that focuses the rays very effectively onto the sample.

Studies are underway to establish whether the transition from elliptic to parabolic geometry should always be around 80\% of the way down the guide system.  Competing with the coma correction from the parabola is the fact that the last half of the ellipse close to the target offers a smaller coma size than the source.  It should be noted that expanding this logic and having a parabolic-elliptic hybrid (the inversion of the current geometry) does not perform as well as the elliptic-parabolic hybrid described here.

The hybrid guide geometry should be applicable in a wide area of neutron instrumentation.  We are particularly interested in the effect upon time-of-flight and diffraction applications, for which the beam homogeneity provides a relatively simple angular resolution function, and also a simple convolution with chopper openings.  Both of these scenarios are likely to be served better by a hybrid system than a pure elliptic system.

An example where hybrid guides should excel is in backscattering spectrometers of the IN16 type \cite{FRICK_IN16}, where the beam divergence distribution function maps onto the instrument dynamic range, and any inhomogeneities in the beam divergence (such as the minima in figures \ref{fig:ballisticDiv} and \ref{fig:ellipseDiv}) directly affect the quality of the instrument's data.

\section{Conclusions}
We have shown that the coma effect in elliptic neutron guides can be greatly reduced by using a hybrid geometry where the second section of the guide is parabolic.  Such hybrid elliptic-parabolic guides are expected to be of interest in a wide range of applications, either as a primary beam delivery system or as a means of focusing a moderately diverging beam onto a sample or virtual source.

\section{Acknowledgments}
The authors are indebted to all the participants of the ANSTO-NIST instrumentation design workshop that ran for the month of November 2009 at NIST, Gaithersburg MD, USA.  In particular, we thank our host Rob Dimeo, NIST scientists Jeremy Cook, John Copley, and John Barker; and the delegates from ANSTO: Anna Sokolova, Nicolas de Souza, Frank Klose and Oliver Kirstein.  We would also like to thank Rohland G\"ahler (Institut Laue Langevin), Gary McIntyre (ANSTO) for useful discussions.  Mention of specific commercial products is for informational purposes only and does not constitute endorsement by the National Institute of Standards and Technology.  This work was presented at International Workshop on Neutron Optics ``NOP2010'' in Alpe d'Huez, France, and the delegates are thanked for their constructive feedback at this event.

\bibliographystyle{elsarticle-num}
\bibliography{ellipticAbberations}

\end{document}